\documentclass[12pt]{iopart}
\usepackage{graphicx}% Include figure files
\usepackage{color}

\newcommand{\Name}[1]{#1}
\newcommand{\REVIEW}[4]{#3 \textit{#1} \textbf{#2} #4}

\begin{document}

\title[Fermi surface and electronic correlations in Sm$_{2-x}$Ce$_{x}$CuO$_4$]
{The Fermi surface and the role of electronic correlations in Sm$_{2-x}$Ce$_{x}$CuO$_4$}

%\rapid{Fermi surface and role of electronic correlations in Sm$_{2-x}$Ce$_{x}$CuO$_4$}

\author{M M Korshunov$^{1,2}$\footnote{Present address:
Department of Physics, University of Florida, Gainesville, Florida 32611, USA},
E V Zakharova$^{2}$, I A Nekrasov$^{3}$, Z V Pchelkina$^{4}$ and S G
Ovchinnikov$^{2,5}$}

\address{$^{1}$  Max-Planck-Institut f\"{u}r Physik komplexer Systeme, D-01187 Dresden,
Germany}
\address{$^{2}$ L.V. Kirensky Institute of Physics, Siberian Branch of Russian
Academy of Sciences, 660036 Krasnoyarsk, Russia}
\address{$^{3}$ Institute for Electrophysics, Russian Academy of Sciences,
Ekaterinburg 620016, Russia}
\address{$^{4}$ Institute for Metal Physics, Russian Academy of Sciences, Ekaterinburg 620219,
Russia}
\address{$^{5}$ Siberian Federal University, 660041 Krasnoyarsk, Russia}

\ead{korshunov@phys.ufl.edu}

\begin{abstract}
Using LDA+GTB (local density approximation+generalized tight-binding) hybrid
scheme we investigate the band structure of the electron-doped high-$T_c$
material Sm$_{2-x}$Ce$_{x}$CuO$_4$. Parameters of the minimal tight-binding
model for this system (the so-called 3-band Emery model) were obtained within
the NMTO ($N$-th order Muffin-Tin orbital) method. Doping evolution of the
dispersion and Fermi surface in the presence of electronic correlations was
investigated in two regimes of magnetic order: short-range (spin-liquid) and
long-range (antiferromagnetic metal). Each regime is characterized by the
specific topologies of the Fermi surfaces and we discuss their relation to
recent experimental data.
%Each regime is characterized by the specific topology of the Fermi surface and
%the comparison to the existing experimental data allows us to claim that the
%optimally doped Sm$_{2-x}$Ce$_{x}$CuO$_4$ exhibits magnetic order close to the
%long-range one.
\end{abstract}

% Cuprate superconductors (high-$T_c$ and insulating parent compounds)
% Electronic structure
% Strongly correlated electron systems: generalized tight-binding method
\submitto{\JPCM}
%\maketitle

\section{Introduction}

One of the most important questions in condensed matter is how the strong
interaction between quasiparticles modify their properties and influence
observable quantities. Non-Fermi-liquid behavior was found in many different
substances, but a class of high-$T_c$ copper oxides attracts special attention
during the last few decades. The unconventional, non-$s$-wave,
superconductivity has a lot to do with it. While other players came to stage,
like lamellar sodium cobalt oxides and Iron-based pnictide superconductors,
only high-$T_c$ cuprates combine both strong electronic correlations and high
values of critical temperature.

A key issue in a theory of high-$T_c$ superconductivity is the proper
description of the low-energy electronic structure. Recent experimental
results, mainly of angle-resolved photoemission spectroscopy (ARPES)
\cite{Yoshida2003,Damascelli2003,Meng2009} and measurements of quantum
oscillations \cite{DoironLeyraud2007,Yelland2008,Sebastian2008}, provide a
pattern to test various theoretical models and schemes. One of the approaches,
proposed by some of the present authors, is the LDA+GTB hybrid scheme
\cite{Korshunov2005}. It was shown that the mean-filed theory within this
scheme captures the most essential features of the doping-dependent evolution
of the quasiparticle band structure and the Fermi surface
\cite{Korshunov2007a,Korshunov2007b}.

A lot of theoretical and experimental efforts were concentrated on the hole
doped compounds. Systems with electron doping, Re$_{2-x}$Ce$_{x}$CuO$_4$
(Re=Nd, Pr, Sm), present a counterpart to hole doped ones and a test for
electron-hole asymmetry in Mott-Hubbard insulators. Recent ARPES data on the
optimally doped, $x=0.14$, Sm-based compound provide detailed information on
the Fermi surface and the band dispersion in the vicinity of the Fermi level
\cite{Park2007}. Similar study was reported for Nd-based compound by Schmitt
\textit{et al.} \cite{Schmitt2008}. These results were confirmed independently
by the measurement of the quantum oscillations \cite{Helm2009}. Also, Park
\textit{et al.} \cite{Park2007} presented an explanation of the observed data
based on the $\sqrt{2} \times \sqrt{2}$ spin-density wave (SDW) model. On the
other hand, the high-energy electronic structure is found to be inconsistent
with the SDW scenario. Moreover, the SDW model implies the weak or moderate
electronic correlations and a Fermi liquid background, which is obviously not
the case for the underdoped and optimally doped cuprates. Thus it is not a
satisfactory scenario and a strong correlation effects should be taken into
account.

Here we present the investigation of electronic structure for the
electron-doped high-$T_c$ material Sm$_{2-x}$Ce$_{x}$CuO$_4$ by means of
LDA+GTB hybrid scheme. Parameters of a minimal generic tight-binding model for
these systems (the so-called Emery model) were obtained within the NMTO method.
Doping evolution of the band structure and the Fermi surface within this model
in presence of strong electronic correlations and magnetic fluctuations were
studied in the framework of the GTB method.

The LDA+GTB electronic structure strongly depends on the underlying magnetic
order. Though there is no N\'eel temperature in the optimally doped n-type
cuprates, the antiferromagnetic (AFM) correlation length is extremely large
($\lambda \approx 400 a$) up to $x=0.17$
\cite{Sumarlin1992,Gukasov1995,Chang2002,Motoyama2007}. Such correlation length
makes magnetic behavior to be rather close to the long-range ordered AFM. That
is why our spin-liquid description is unable to capture some details of the
observed Fermi surface. On the other hand, there is a good agreement with the
recent ARPES data once we assume the presence of the long-range order.

\section{Noninteracting band structure \label{results_LDA}}

Here we will describe the noninteracting band structure and in the next Section
introduce the electronic correlations within the LDA+GTB method.

Sm$_2$CuO$_4$ system has body-centered tetragonal crystal structure with the
space group $I4/mmm$. Values of lattice parameters are $a=3.917$\AA and
$c=11.899$\AA. The atomic positions for different atoms are: Cu (0,0,0), Sm
(0,0,0.35184) and two types of oxygens O1 (0,0.5,0), O2
(0,0.5,0.25)~\cite{struct}. Physically important CuO$_2$ layers are constructed
with O1 type oxygens. No apical oxygen is presented in this structure.

\begin{figure}
\begin{center}
\includegraphics[clip=true,width=0.7\columnwidth]{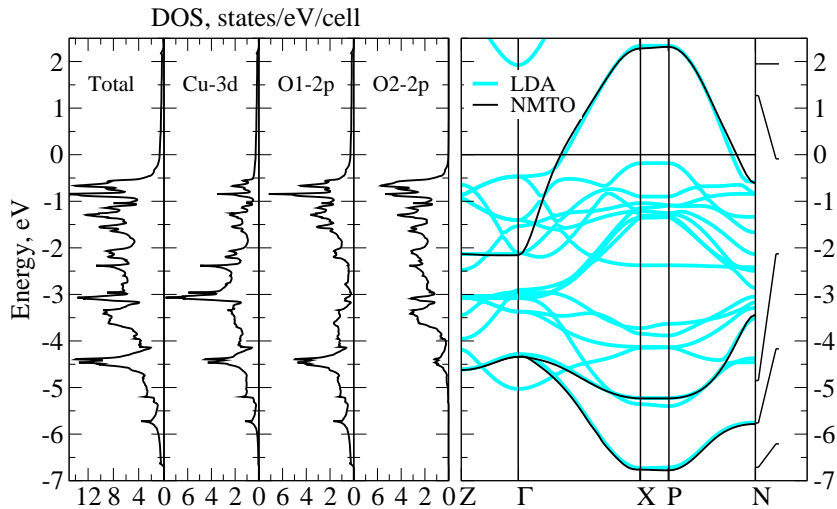}
\caption{Electronic structure of Sm$_2$CuO$_4$ obtained within LDA.
Left panel presents total and partial densities of states; right panel shows
band dispersions: thick (cyan) curves denote LMTO bands,
while thin (black) curves denote NMTO bands.
Zero corresponds to the Fermi level.}
\label{dos_band}
\end{center}
\end{figure}

We perform density functional theory band structure calculations within linear
muffin-tin orbital basis set employing atomic sphere approximation in the
framework of program package TB-LMTO-ASA v47~\cite{LMTO1,LMTO2,LMTO3}. In
Fig.~\ref{dos_band} results of our LMTO computations are presented. Left panel
shows the total and the partial densities of states. Cu-3$d$ and O1-2$p$ states
cross the Fermi level. In the right panel of Fig.~\ref{dos_band} LMTO band
dispersions are presented (thick curves). Note the Fermi level is crossed by
just one antibonding hybrid Cu-3$d$---O1-2$p$ band of $x^2-y^2$ symmetry. It is
in agreement with the generic minimal tight-binding model for high-$T_c$
cuprates \cite{emery1,Varma}. Orbital basis for this model consists of
Cu-$3d_{x^2-y^2}$ orbital and in-plane $p_x$ and $p_y$ oxygen orbitals. To
compute corresponding model parameters $N$-th order Muffin-Tin orbital method
(NMTO)~\cite{oka2000} was used. Necessary for NMTO expansion energies are
schematically shown on the right side of Fig.~\ref{dos_band}. Obtained hopping
parameters are listed in the Table~\ref{tab1}, and the single electron energies
are E$_{x^2-y^2}=-2.322$ eV and E$_{p_x}=-3.708$ eV.

We assume that the values of Coulomb repulsion $U$ and Hund's exchange $J_H$
for Cu ions are doping independent and equal to 10 eV and 1 eV, respectively
(see Ref.~\cite{Korshunov2005} for details).
\begin{table}[t]
\caption{\label{tab1} Parameters for the 3-band model obtained within NMTO
method. Here $x^2$ and $p_{x,y}$ denote the Cu-$3d_{x^2-y^2}$ and O-$p_{x,y}$
orbital indexes. All values are in eV.}
\begin{indented}
\item[] \begin{tabular}{@{}lllllll} \br
hopping&\centre{6}{involved orbitals}\\
\mr
&\centre{2}{($x^2$,$p_x$)}&\centre{2}{($x^2$,$x^2$)}&\centre{2}{($p_x$,$p_y$)}\\
\ns
&\crule{2}&\crule{2}&\crule{2}\\
& direction & value & direction & value & direction & value \\
\ns
&\crule{2}&\crule{2}&\crule{2}\\
t        & (0.5,0) & 1.261 & (1,0)  & 0.138 & (0.5,0.5) & 0.882 \\
t$'$     & (0.5,1) &-0.011 & (1,1)  &-0.025 & (1.5,0.5) & 0.033 \\
t$''$    & (1.5,0) & 0.1   & (2,0)  & 0.011 & (1.5,1.5) & 0.021 \\
t$'''$   & (1.5,1) &-0.007 & (1,2)  &-0.012 & (2.5,0.5) & 0.005 \\
%t$^{IV}$ & (0.5,2) & 0.003 &        &       &           &       \\
\br
\end{tabular}
\end{indented}
\end{table}

\section{LDA+GTB scheme}

Within the LDA+GTB method \cite{Korshunov2005} the results of \textit{ab
initio} band structure calculations presented in the previous Section are used
to construct the Wannier functions and to obtain the parameters of the
multiband Hubbard-type model. For this multiband model the electronic structure
in the strong correlation regime is calculated within the generalized
tight-binding (GTB) method
\cite{Ovchinnikov1989,Gavrichkov2000,Gavrichkov2001}. The latter combines the
exact diagonalization of the model Hamiltonian for a small cluster (unit cell)
with perturbative treatment of the intercluster hopping and interactions. After
this step we end up with the GTB Hamiltonian. Depending on the order of
perturbative treatment, we can formulate different approximations.

As was shown before \cite{Korshunov2005}, for undoped and weakly doped
La$_{2-x}$Sr$_x$CuO$_4$ and Nd$_{2-x}$Ce$_x$CuO$_4$ this scheme in the lowest
order in hopping (Hubbard-I approximation) results in a charge transfer
insulator with a correct value of the gap $E_{ct}$ and the dispersion of bands
in agreement with the experimental ARPES data.

We map the GTB Hamiltonian onto the effective $t-t'-t''-J^*$ model, where
``$*$'' denotes the three-site correlated hoppings, and study two regimes of
AFM correlations. Namely, the spin-liquid phase with short-range AFM
fluctuations and the long-range AFM metallic phase. We will describe AFM
metallic phase using the Hubbard-I approximation that was shown to be in
qualitative agreement with the Quantum Monte-Carlo results
\cite{Ovchinnikov2003}. To study the spin-liquid phase, we use the same
procedure as in Ref.~\cite{Korshunov2007a} and go beyond the Hubbard-I
approximation: (i) We solve the Dyson equation in the paramagnetic phase by
means of the diagram technique for the Hubbard $X$-operators
\cite{Ovchinnikov_book2004} and (ii) Obtain the coupled equations for the
self-energy $\hat{\Sigma}(\mathbf{k},\omega)$, the strength operator
$\hat{P}(\mathbf{k},\omega)$, and the spin-spin and kinematic correlation
functions, (iii) We solve the coupled equations self-consistently and obtain a
doping dependent Fermi surface and band structure.

The $t-t'-t''-J^*$ model Hamiltonian is given by
\begin{eqnarray}
H_{t-J^*}&=& H_{t - J} + H_{(3)}, \\
H_{t-J} &=& \sum\limits_{f,\sigma } (\varepsilon - \mu )X_f^{\sigma \sigma}
+ \sum\limits_{f \ne g,\sigma } t_{fg} X_f^{\sigma 0} X_g^{0 \sigma}
+ \sum\limits_{f \ne g} {J_{fg} \left( {\mathbf{S}_f \cdot \mathbf{S}_g - \frac{1}{4} n_f n_g } \right)} \nonumber\\
H_{(3)} &=& \sum\limits_{f \ne m \ne g,\sigma } {\frac{\tilde{t}_{fm} \tilde{t}_{mg}}{U}
\left( {X_f^{\sigma 0} X_m^{\bar\sigma \sigma } X_g^{0 \bar\sigma } -
X_f^{\sigma 0} X_m^{\bar\sigma \bar\sigma } X_g^{0 \sigma}} \right)}, \nonumber
\end{eqnarray}
where $X_f^{n n'} \equiv \left| n \right> \left< n' \right|$ are the Hubbard
$X$-operators \cite{Hubbard1964} acting on the Hilbert space of local states
$\left| n \right> = \left\{ 0, \sigma, -\sigma \equiv \bar\sigma \right\}$,
$J_{fg} = 2 \tilde{t}_{fg}^2/U$ is the AFM exchange between two sites $f$ and
$g$, $U=E_{ct}$ is the effective Hubbard repulsion determined by the charge
transfer energy $E_{ct} \approx 2$eV, $t_\mathbf{k}=2 t (\cos{k_x}+\cos{k_y}) +
4 t' \cos{k_x}\cos{k_y} + 2 t'' (\cos{2k_x}+\cos{2k_y})$ is the Fourier
transform of the hopping $t_{fg}$, and $\tilde{t}_\mathbf{k}=2 \tilde{t}
(\cos{k_x}+\cos{k_y}) + 4 \tilde{t}' \cos{k_x}\cos{k_y} + 2 \tilde{t}''
(\cos{2k_x}+\cos{2k_y})$ is the Fourier transform of the interband hopping
parameter $\tilde{t}_{fg}$, $\mathbf{S}_f$ is the spin operator, $\varepsilon$
is the one-hole local energy, and $\mu$ is the chemical potential. The Green
function in terms of the Hubbard $X$-operators is
\begin{equation}
\label{eq:G}
G_\sigma(\mathbf{k},\omega) = \left\langle \left\langle \left. X_\mathbf{k}^{0 \sigma}
\right| X_\mathbf{k}^{\sigma 0} \right\rangle \right\rangle_\omega = \frac{(1+x)/2}
{\omega - \varepsilon _0 + \mu - \frac{1 + x}{2} t_{\mathbf{k}} - \frac{1 - x^2}{4}
\frac{\tilde{t}_\mathbf{k}^2}{U} - \Sigma(\mathbf{k})}.
\end{equation}
Within our approximations \cite{Korshunov2007a}, the strength operator
$\hat{P}(\vec{k},E)$ is replaced by the occupation factor $(1+x)/2$ and the
self-energy $\hat{\Sigma}(\mathbf{k},\omega)$ is frequency independent but
preserve momentum dependence,
\begin{eqnarray}
\Sigma(\mathbf{k}) &=& \frac{2}{1 + x}\frac{1}{N}
 \sum\limits_\mathbf{q} \left\{ \left[ t_\mathbf{q} - \frac{1 - x}{2}J_{\mathbf{k} - \mathbf{q}}
 - x \frac{\tilde{t}_\mathbf{q}^2}{U} - (1 + x) \frac{\tilde{t}_\mathbf{k} \tilde{t}_\mathbf{q}}{U} \right]
 K_\mathbf{q} \right. \\
&+& \left. \left[ t_{\mathbf{k} - \mathbf{q}} - \frac{1 - x}{2}\left(J_\mathbf{q}
 - \frac{\tilde{t}_{\mathbf{k} - \mathbf{q}}^2}{U} \right)
 - (1 + x) \frac{\tilde{t}_\mathbf{k} \tilde{t}_{\mathbf{k} - \mathbf{q}}}{U} \right]
 \cdot \frac{3}{2} C_\mathbf{q} \right\}.
\end{eqnarray}
The spin-spin $C_{\textbf{q}}$ and kinematic $K_{\textbf{q}}$ correlation
functions play significant role representing the short-range AFM fluctuations
and the kinetic energy reduced by the correlation effects, respectively:
\begin{eqnarray}
\label{corr_func}
C_\mathbf{q} &=& \sum\limits_{\mathbf{f} - \mathbf{g}} e^{-\mathrm{i}(\mathbf{f} - \mathbf{g})\mathbf{q}}
 \left\langle X_\mathbf{f}^{\sigma \bar\sigma} X_\mathbf{g}^{\bar\sigma \sigma} \right\rangle
= 2\sum\limits_{\mathbf{f} - \mathbf{g}} e^{-\mathrm{i}(\mathbf{f} - \mathbf{g})\mathbf{q}}
 \left\langle S_\mathbf{f}^z S_\mathbf{g}^z \right\rangle, \nonumber\\
K_\mathbf{q} &=& \sum\limits_{\mathbf{f} - \mathbf{g}} e^{-\mathrm{i}(\mathbf{f} - \mathbf{g})\mathbf{q}}
 \left\langle X_\mathbf{f}^{\sigma 0} X_\mathbf{g}^{0 \sigma} \right\rangle.
\end{eqnarray}

Energy spectrum is determined by the poles of the Green function (\ref{eq:G})
and Fermi surface is determined by the equation $\varepsilon_0 - \mu + \frac{1
+ x}{2} t_{\mathbf{k}} + \frac{1 - x^2}{4} \frac{\tilde{t}_\mathbf{k}^2}{U} +
\Sigma(\mathbf{k})$ = 0.

\section{Results and discussion}

The procedure of mapping the GTB Hamiltonian onto the effective model was
described in detail in Ref.~\cite{Korshunov2005}. Following the same steps and
using the parameters listed in Table~\ref{tab1}, we obtain the $t-t'-t''-J^*$
model with the following hoppings and exchange interactions: $t=-0.59$ eV,
$t'=-0.08 t$, $t''=0.15 t$, $J=0.92 |t|$, $J'=0.01 |t|$, $J''=0.02 |t|$,
$\tilde{t}=-0.74$ eV, $\tilde{t}'=-0.11 \tilde{t}$, $\tilde{t}''=0.16
\tilde{t}$. Here, $\tilde{t}$, $\tilde{t}'$, and $\tilde{t}''$ are the
interband hoppings through the charge-transfer gap, which determine the
three-site hoppings and the exchange parameter, $J=2 \tilde{t}^2/E_{ct}$. Note
that although the value of the nearest-neighbor exchange $J$ is quite large,
the spin gap in the AFM phase will be determined not by this value alone, also
there will be a contribution from the three-site hoppings. This contribution
reduce the value of the spin gap as will be discussed later.

\begin{figure}
\begin{center}
\includegraphics[clip=true,angle=0,width=0.7\columnwidth]{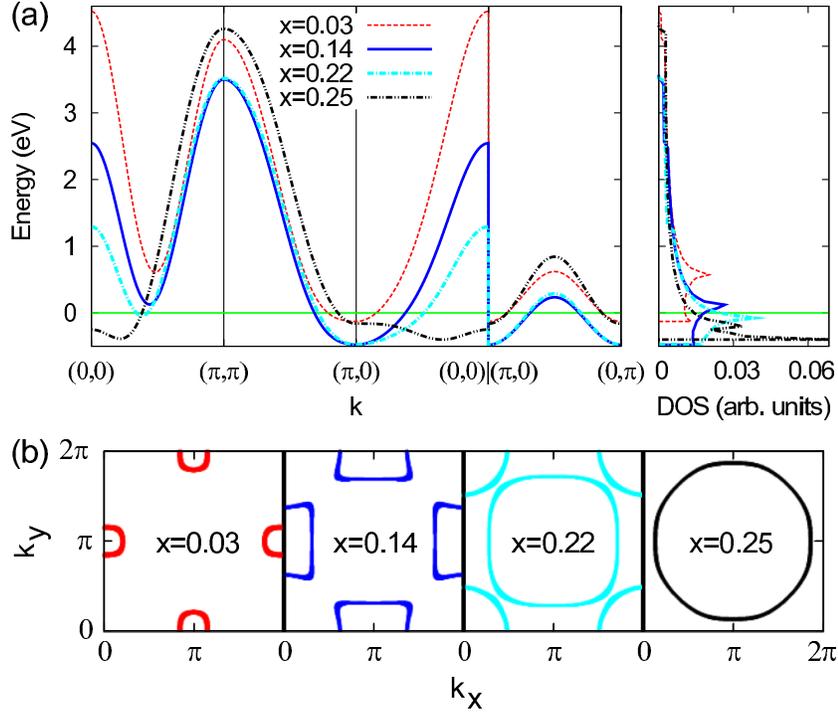}
\caption{Spin-liquid phase: Band structure and DOS (a) and the Fermi surface (b) for
Sm$_{2-x}$Ce$_{x}$CuO$_4$ within LDA+GTB method for different doping concentrations $x$, as indicated.
In (a) zero corresponds to the Fermi level.}
\label{spinc}
\end{center}
\end{figure}
In Fig.~\ref{spinc} we present results for $t-t'-t''-J^*$ model in the
spin-liquid phase. At low doping, $x=0.03$, due to the scattering on the
magnetic fluctuations, the band structure possess local AFM symmetry in the
vicinity of the $(\pm\pi/2,\pm\pi/2)$ points [see Fig.~\ref{spinc}(a)] and the
Fermi surface has a form of four electron pockets around the $(0,\pm\pi)$ and
$(\pm\pi,0)$ points. Values of the spin-spin correlation functions
$C_{\textbf{q}}$ are large enough for the similar topology to survive until $x
\approx 0.22$, where a quantum phase transition with change of the Fermi
surface topology takes place. After the transition, the Fermi surface at
$x=0.22$ has a form of a large hole pocket around the $(\pi,\pi)$ point and a
small hole pocket around the $(0,0)$ point, which decrease in size with further
electron doping. At $x=0.25$ only one large hole pocket around the $(\pi,\pi)$
point is left. The quantum phase transition with the change of the Fermi
surface topology was found experimentally in Nd-based compound \cite{Helm2009},
though at a different critical concentration.

Note that the standard formulation of the Luttinger theorem does not work for
the Hubbard fermions since the spectral weight of such fermion is determined by
the strength operator, $\hat{P}(\textbf{k},E) = F_{0 \sigma}$, and each quantum
state contains $2 F_{0 \sigma} = 1 - x$ electrons. A generalized Luttinger
theorem for the strongly correlated systems~\cite{Korshunov2003} takes into
account the spectral weight of each $\left| k \right\rangle$ state and the
Fermi surface in our Fig.~\ref{spinc} satisfies its completely.

\begin{figure}
\begin{center}
\includegraphics[clip=true,angle=0,width=0.45\columnwidth]{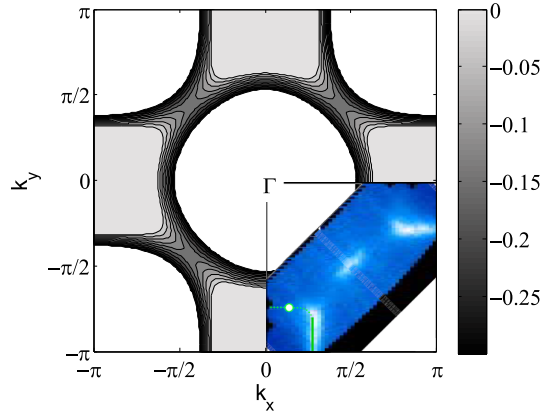}
\caption{Fermi surface ``mapping'' (set of equipotential cuts from the Fermi level to -0.3 eV below it)
in the spin-liquid phase for $x=0.14$ together with the reproduction of experimental ARPES data
from Ref.~\cite{Park2007} in the lower right corner.
Colorbar on the right shows the correspondence between shades of gray
and energy from the Fermi level (in eV).}
\label{fsmap}
\end{center}
\end{figure}

A comparison of the calculated Fermi surface in the spin-liquid phase and the
experimental ARPES data \cite{Park2007} is shown in Fig.~\ref{fsmap}. Note the
difference in the methods to obtain the Fermi surface ``mapping''. We draw a
set of constant energy cuts from the Fermi level down to -0.3 eV below it,
while the experimental Fermi surface mapping is an integration of ARPES
intensities over 30 meV energy window. The ARPES Fermi surface consists of
three parts: two pockets around $(0,\pi)$ and $(\pi,0)$ points, and one
elongated pocket around $(\pi/2,\pi/2)$ point. One can immediately notice from
Fig.~\ref{spinc}(b) that in our spin-liquid theory the pocket around
$(\pi/2,\pi/2)$ point is missing; it does not appear even if one collects
intensities from below the Fermi level, as seen in Fig.~\ref{fsmap}. Moreover,
there are no features in the band dispersion, which could produce such pocket.
Thus we conclude that our theory for the spin-liquid phase does not reproduce
all details of the experimental Fermi surface.

Since optimally doped Sm$_{2-x}$Ce$_{x}$CuO$_4$ is in the vicinity of the
ordered AFM phase and the correlation length is extremely large (about 400
lattice constants) \cite{Motoyama2007}, we now investigate the band structure
in the $t-t'-t''-J^*$ model assuming the long-range AFM order. The procedure is
similar to Refs.~\cite{Ovchinnikov2004,Ovchinnikov2008}, where the energy
spectrum of the $t-t'-t''-J$ model was obtained within the Hubbard-I
approximation, but here we also take the three-site hopping terms into account.

\begin{figure}
\begin{center}
\includegraphics[clip=false,angle=0,width=0.7\columnwidth]{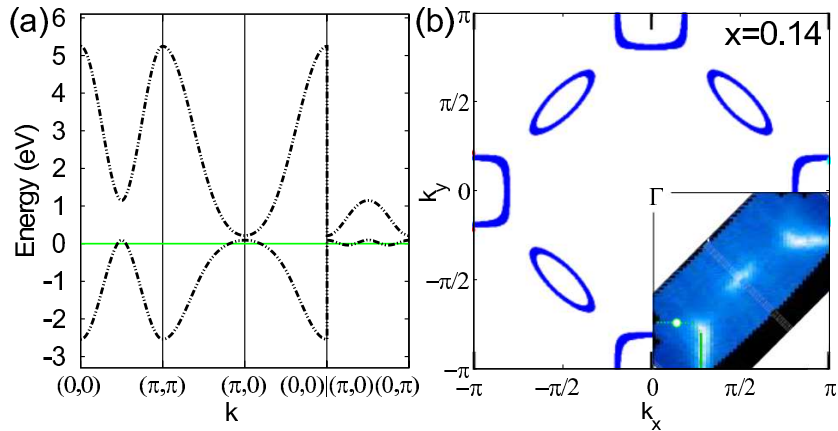}
\caption{AFM phase: Band structure (a) and the Fermi surface (b) for $x=0.14$ within LDA+GTB method.
In the lower right corner of (b) we show the reproduction of the ARPES
Fermi surface map from Ref.~\cite{Park2007}.
In (a) zero corresponds to the Fermi level.}
\label{afm3c}
\end{center}
\end{figure}

In Fig.~\ref{afm3c}(b) we present results for the Fermi surface in the AFM
phase of the $t-t'-t''-J^*$ model at $x=0.14$ together with the experimental
ARPES data. Evidently, there is a rather good agreement between both. We would
like to mention that for lower concentrations our calculations result in
decrease of pockets around the $(\pi/2,\pi/2)$ points and increase of pockets
around the $(\pi,0)$ and $(0,\pi)$ points. Note that in the $t-J$ model the
spin gap is determined solely by the AFM exchange $J$ \cite{Ovchinnikov2004}.
Here, momentum dependence of the spin gap is proportional to $t'\cos{k_x}
\cos{k_y}$ (see Eq.~(12) of Ref.~\cite{Ovchinnikov2004}). The reason is that in
the absence of spin fluctuations the hoping of a particle without spin flip
processes possible only within the same spin sublattice. Because of the
$\cos{k_x} \cos{k_y}$ functional form the spin gap is maximal at
$(\pi/2,\pi/2)$ point and minimal at $(\pi,0)$ point as seen in
Fig.~\ref{afm3c}(a). Since the three-site hopping terms involving sites $f$,
$m$, and $g$ are proportional to $\tilde{t}_{fm} \tilde{t}_{mg}/E_{ct}$, they
also contribute to the spin gap; but, apparently, they decrease the gap value
around $(\pi,0)$ point making it more anisotropic.

Since we are making a mean-filed theory (though in a strong interaction limit)
we can not address the question of the intensity distribution over the Fermi
surface. This question was addressed earlier by different groups
\cite{Aichhorn2006,Kokorina2008}. Remarkably, their results on the Fermi
surface contours for $x \approx 0.14$ are very similar to our's in
Fig.~\ref{afm3c} in spite of rather different calculation schemes. This again
emphasizes the fact that the AFM correlations are very strong in the optimally
electron doped cuprates and they determine the quasiparticle dispersion and the
Fermi surface.

\section{Conclusion}

We have shown that the experimentally observed Fermi surface topology can be
explained within the LDA+GTB calculations for the long-range AFM spin
background. On the other hand, our theory for the spin-liquid phase
demonstrates only partial agreement with the ARPES Fermi surface due to the
underestimation of the impact of magnetic scattering on the electronic
structure. We conclude that the spin fluctuations are very strong in
Sm$_{1.86}$Ce$_{0.14}$CuO$_4$ and are closer to the long-range AFM fluctuations
rather than to the fluctuations in the spin-liquid phase. Similar conclusion
was drawn recently from the analysis of quantum oscillations in Nd-based
electron doped cuprates \cite{Helm2009}.

We would like to emphasize the significant difference between our picture for
AFM order and one by Park \textit{et al.} \cite{Park2007}. Park \textit{et al.}
provide a simple calculation based on the conventional SDW order \textit{i.e.}
the one based on a weak coupling approximation for the interaction. In the
absence of the long-range order the ground state is metallic even at zero
doping, $x=0$. On the other hand, our approach allows to study the limit of
large interaction and provides an insulating ground state at zero doping. This
is essential difference since the underdoped cuprates belong to a class of
strongly interacting systems and exhibit a Mott transition at a half filling,
$x=0$. More precisely, because of the copper-oxygen hybridization the cuprates
shows the charge-transfer gap $E_{ct}$ at $x=0$, but on the level of a
single-band Hubbard model one can speak about a Mott-Hubbard effective gap
$U=E_{ct}$.

\ack
We would like to thank A.A. Kordyuk and I. Eremin for useful discussions. The
authors acknowledge support from RFBR (grants 08-02-00021, 10-02-00662,
08-02-91200, 07-02-00226), RAS programs on ``Low temperature quantum
phenomena'', ``Quantum physics of condensed matter'' and ``Strongly correlated
electrons solids''. President of Russia [grants MK-614.2009.2 (I.N.) and
MK-3227.2008.2 (Z.P.)], scientific school (grant SS-1929.2008.2),
interdisciplinary UB-SB RAS project, Dynasty Foundation (Z.P.), and Russian
Science Support Foundation (I.N.).

\end{document}